%% file: main.tex
\documentclass[letterpaper]{article} % DO NOT CHANGE THIS
\usepackage{aaai24}  % DO NOT CHANGE THIS
\usepackage{times}  % DO NOT CHANGE THIS
\usepackage{helvet}  % DO NOT CHANGE THIS
\usepackage{courier}  % DO NOT CHANGE THIS
\usepackage[hyphens]{url}  % DO NOT CHANGE THIS
\usepackage{graphicx} % DO NOT CHANGE THIS
\usepackage{natbib}  % DO NOT CHANGE THIS AND DO NOT ADD ANY OPTIONS TO IT
\usepackage{caption} % DO NOT CHANGE THIS AND DO NOT ADD ANY OPTIONS TO IT
\usepackage{calc}
\usepackage{amsmath}
\usepackage{caption}
\usepackage{multirow}
\usepackage{amsfonts}
\usepackage[font=footnotesize]{subcaption}
\usepackage{booktabs}
\usepackage{pdfpages}
\usepackage{xcolor}
\usepackage{tabularx}
\usepackage{algorithm}
\usepackage{algorithmic}

\newcommand{\kiss}{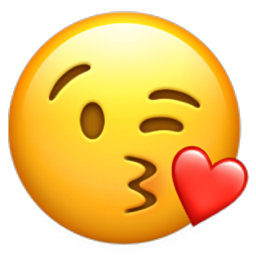}
\newcommand{\sparkles}{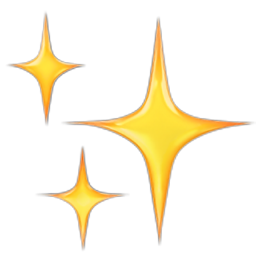}

\newcommand{\pleadingface}{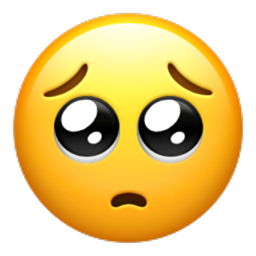}
\newcommand{\heart}{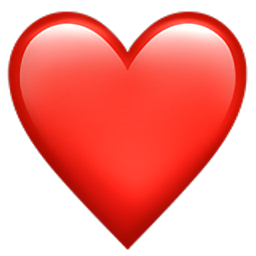}

\newcommand{\fire}{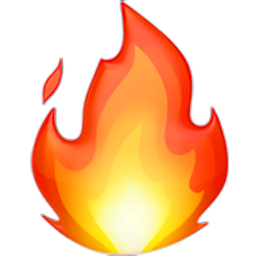}

\newcommand{\pray}{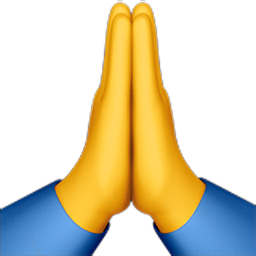}

\newcommand{\joy}{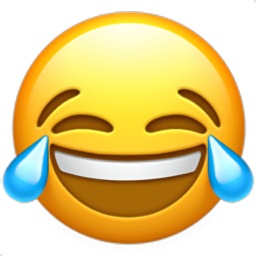}

\newcommand{\sob}{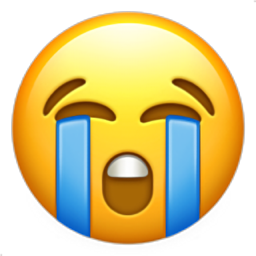}
\newcommand{\hearteyes}{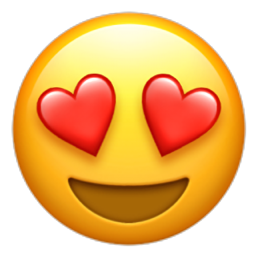}

\newcommand{\twohearts}{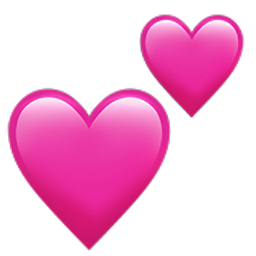}

\newcommand{\wink}{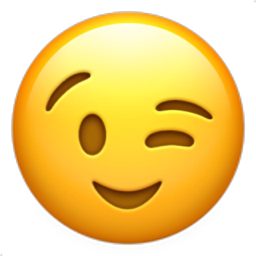}

\newcommand{\pensiveface}{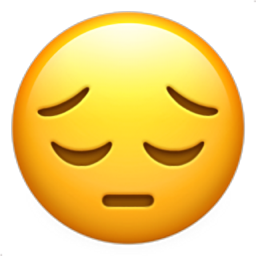}
\newcommand{\checkmarkemoji}{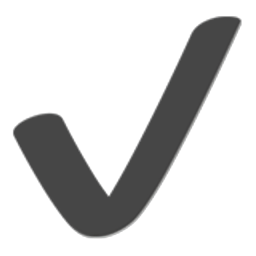}
\newcommand{\crossmarkemoji}{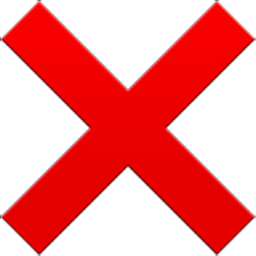}
\newcommand{\thumbsup}{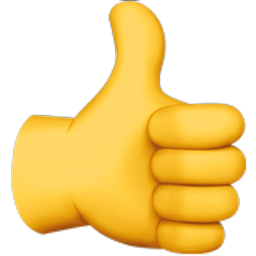}
\newcommand{\stopsign}{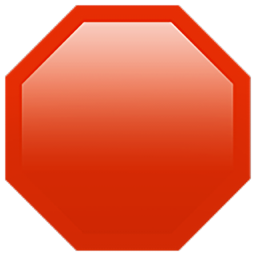}
\newcommand{\smilingeyes}{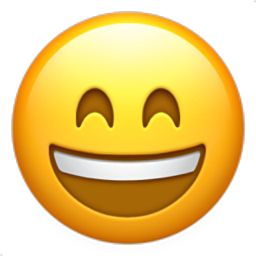}
\newcommand{\rocket}{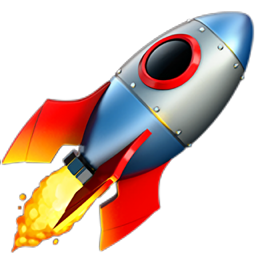}
\newcommand{\wavinghand}{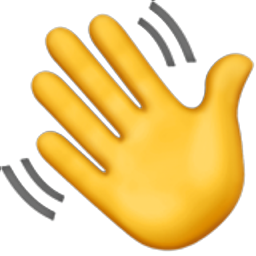}
\newcommand{\partypopper}{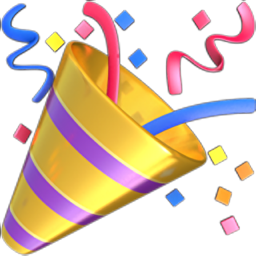}
\newcommand{\whitecheckmark}{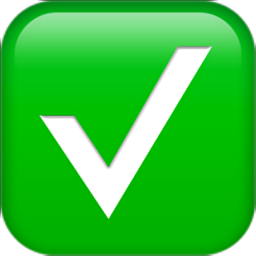}
\newcommand{\bugemoji}{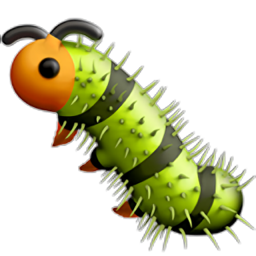}
\newcommand{\warningemoji}{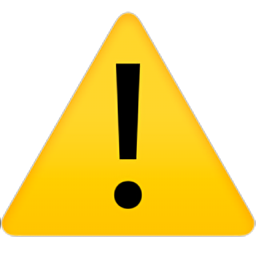}
\newcommand{\thinkingface}{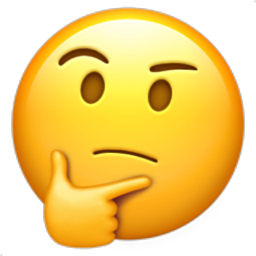}
\newcommand{\slightsmilingface}{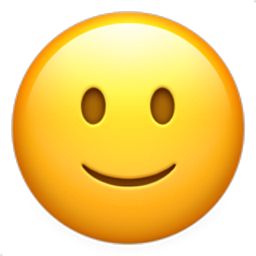}
\newcommand{\grinningface}{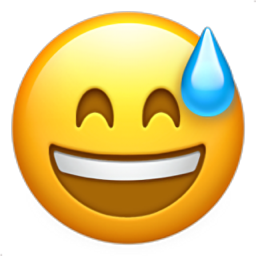}
\newcommand{\ladybugemoji}{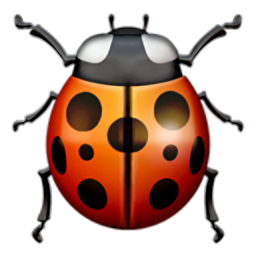}
\newcommand{\smilingfacesmilingeyes}{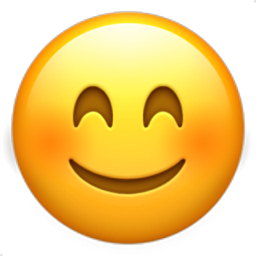}
\newcommand{\moneybag}{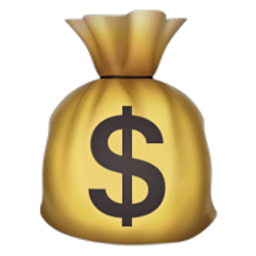}
\newcommand{\triangular}{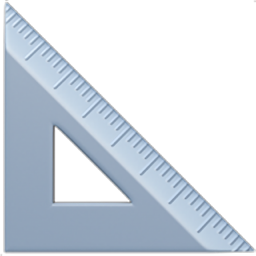}
\newcommand{\bigeyes}{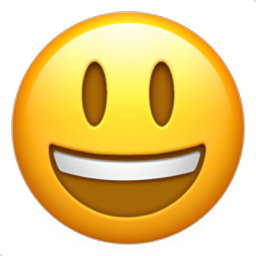}
\newcommand{\heartsuitemoji}{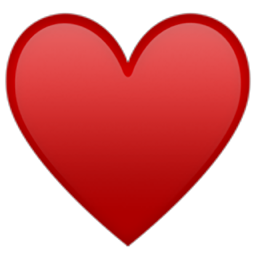}
\newcommand{\rollinglaughing}{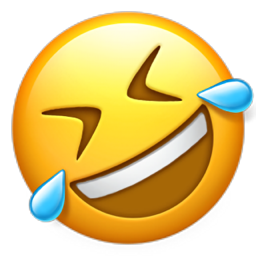}
\newcommand{\smilingfacewithhearts}{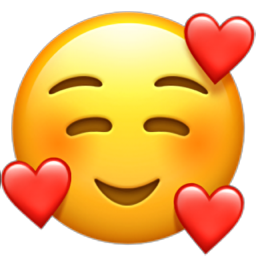}
\newcommand{\beamingface}{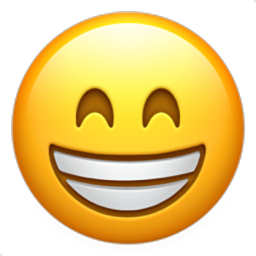}
\newcommand{\smilingface}{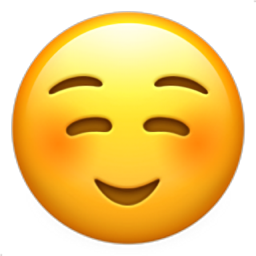}
\newcommand{\facepalming}{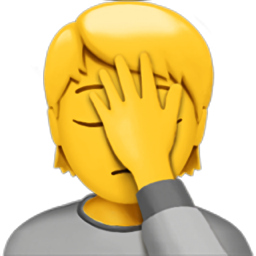}
\newcommand{\shrugging}{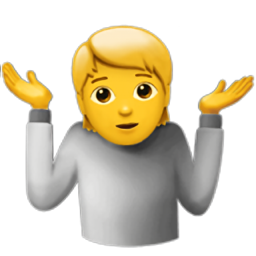}
\newcommand{\clipboard}{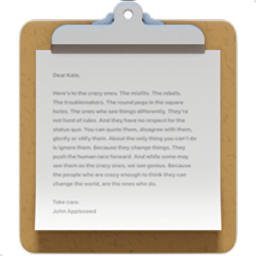}
\newcommand{\speechballon}{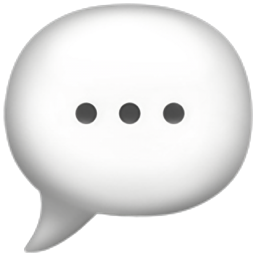}
\newcommand{\openhands}{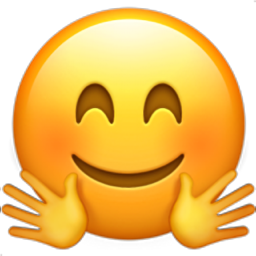}
\newcommand{\robot}{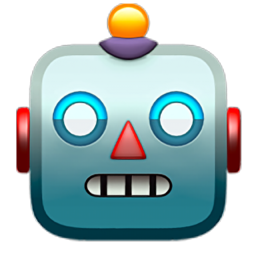}

\newcommand{\goldstar}{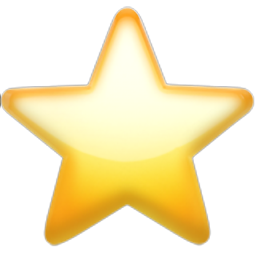}

\newcommand{\facewithmonocle}{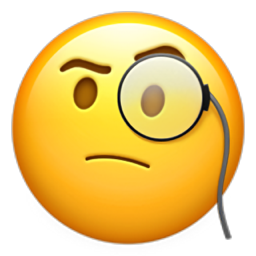}

\usepackage[export]{adjustbox}
\newcommand{\MyEmoji}[1]{\includegraphics[width=1em,valign=t]{#1}}

\newcommand{\answerYes}[1]{\textcolor{blue}{#1}} 
\newcommand{\answerNo}[1]{\textcolor{teal}{#1}} 
\newcommand{\answerNA}[1]{\textcolor{gray}{#1}}

\usepackage{newfloat}
\usepackage{listings}
\floatstyle{ruled}
\newfloat{listing}{tb}{lst}{}
\floatname{listing}{Listing}
\title{Emoji Promotes Developer Participation and Issue Resolution on GitHub}

\author {
    Yuhang Zhou\textsuperscript{1},
    Xuan Lu\textsuperscript{2},
    Ge Gao\textsuperscript{1},
    Qiaozhu Mei\textsuperscript{3},
    Wei Ai\textsuperscript{1} \\
}
\affiliations {
    \textsuperscript{1} College of Information Studies, University of Maryland, College Park, USA \\
    \textsuperscript{2} School of Information, University of Arizona, Tucson, USA \\
    \textsuperscript{3} School of Information, University of Michigan, Ann Arbor, USA \\
    tonyzhou@umd.edu, luxuan@arizona.edu, gegao@umd.edu, qmei@umich.edu, aiwei@umd.edu
}

\begin{document}

\maketitle

\begin{abstract}
Although remote working is increasingly adopted during the pandemic, many are concerned by the low-efficiency of remote working. Missing in text-based communication are non-verbal cues such as facial expressions and body language, which hinders effective communication and negatively impacts work outcomes. Prevalent on social media platforms, emojis, as alternative non-verbal cues, are gaining popularity in the virtual workspaces well. In this paper, we study how emoji usage influences developer participation and issue resolution in virtual workspaces. To this end, we collect GitHub issues for a one-year period and apply causal inference techniques to measure the causal effect of emojis on the outcome of issues, controlling for confounders such as issue content, repository, and author information. We find that emojis can significantly reduce the resolution time of issues and attract more user participation. We also compare the heterogeneous effect on different types of issues. These ﬁndings deepen our understanding of the developer communities, and they provide design implications on how to facilitate interactions and broaden developer participation.
\end{abstract}

\input{sections/1_introduction}

\input{sections/2_related}

\input{sections/3_data}
% \input{sections/4_timeline}
\input{sections/4_confounders}

\input{sections/5_method}

\input{sections/6_result}
\input{sections/7_implication}
\input{sections/8_conclusion}
% \vspace{-0.5em}

% \bibliographystyle{abbrev}
\bibliography{abbrev, aaai24}

\subsection*{Paper Checklist}
\begin{enumerate}

\item For most authors...
\begin{enumerate}
    \item  Would answering this research question advance science without violating social contracts, such as violating privacy norms, perpetuating unfair profiling, exacerbating the socio-economic divide, or implying disrespect to societies or cultures?
    \answerYes{Yes}
  \item Do your main claims in the abstract and introduction accurately reflect the paper's contributions and scope?
    \answerYes{Yes, see Section \textit{Abstract} and \textit{Introduction}}
   \item Do you clarify how the proposed methodological approach is appropriate for the claims made? 
    \answerYes{Yes, see Section \textit{GitHub Issues and Problem Formulation}, \textit{Collection of Confounders} and \textit{Causal Inference of Emojis on Issues}}
   \item Do you clarify what are possible artifacts in the data used, given population-specific distributions?
    \answerNo{No, the GitHub issues do not have demographic information}
  \item Did you describe the limitations of your work?
    \answerYes{Yes, see Section \textit{Limitations and Implications}}
  \item Did you discuss any potential negative societal impacts of your work?
    \answerYes{Yes, see Section \textit{Limitations and Implications}}
      \item Did you discuss any potential misuse of your work?
    \answerNo{No, we do not have a major concern about the potential misuse of our work}
    \item Did you describe steps taken to prevent or mitigate potential negative outcomes of the research, such as data and model documentation, data anonymization, responsible release, access control, and the reproducibility of findings?
    \answerNo{No, we do not have a major concern about the potential misuse}
  \item Have you read the ethics review guidelines and ensured that your paper conforms to them?
    \answerYes{Yes}
\end{enumerate}

\item Additionally, if your study involves hypotheses testing...
\begin{enumerate}
  \item Did you clearly state the assumptions underlying all theoretical results?
    \answerYes{Yes, see Section \textit{GitHub Issues and Problem Formulation}, \textit{Collection of Confounders} and \textit{Causal Inference of Emojis on Issues}}
  \item Have you provided justifications for all theoretical results?
    \answerYes{Yes, see Section \textit{GitHub Issues and Problem Formulation}, \textit{Collection of Confounders}, \textit{Causal Inference of Emojis on Issues} and \textit{Results and Discussion}}
  \item Did you discuss competing hypotheses or theories that might challenge or complement your theoretical results?
    \answerYes{Yes, see Section \textit{GitHub Issues and Problem Formulation}, \textit{Collection of Confounders}, \textit{Causal Inference of Emojis on Issues} and \textit{Results and Discussion}}
  \item Have you considered alternative mechanisms or explanations that might account for the same outcomes observed in your study?
    \answerNo{No}
  \item Did you address potential biases or limitations in your theoretical framework?
    \answerYes{Yes, see Section \textit{Results and Discussion}}
  \item Have you related your theoretical results to the existing literature in social science?
    \answerYes{Yes, see Section \textit{Related Work} and \textit{Limitations and Implications}}
  \item Did you discuss the implications of your theoretical results for policy, practice, or further research in the social science domain?
    \answerYes{Yes, see Section \textit{Limitations and Implications}}
\end{enumerate}

\item Additionally, if you are including theoretical proofs...
\begin{enumerate}
  \item Did you state the full set of assumptions of all theoretical results?
    \answerYes{Yes, see Section \textit{GitHub Issues and Problem Formulation}}
	\item Did you include complete proofs of all theoretical results?
    \answerNA{NA}
\end{enumerate}

\item Additionally, if you ran machine learning experiments...
\begin{enumerate}
  \item Did you include the code, data, and instructions needed to reproduce the main experimental results (either in the supplemental material or as a URL)?
    \answerYes{Yes, see Section \textit{Collection of Confounders}, \textit{Causal Inference of Emojis on Issues} and \textit{Results and Discussion}}
  \item Did you specify all the training details (e.g., data splits, hyperparameters, how they were chosen)?
    \answerYes{Yes, see Section \textit{Collection of Confounders} and \textit{Causal Inference of Emojis on Issues}}
     \item Did you report error bars (e.g., with respect to the random seed after running experiments multiple times)?
    \answerNo{No, the variation is small}
	\item Did you include the total amount of compute and the type of resources used (e.g., type of GPUs, internal cluster, or cloud provider)?
    \answerNo{No}
     \item Do you justify how the proposed evaluation is sufficient and appropriate to the claims made? 
    \answerYes{Yes, see Section \textit{Collection of Confounders} and \textit{Causal Inference of Emojis on Issues}}
     \item Do you discuss what is ``the cost`` of misclassification and fault (in)tolerance?
    \answerYes{Yes, see Section \textit{Collection of Confounders}, \textit{Causal Inference of Emojis on Issues} and \textit{Results and Discussion}}
  
\end{enumerate}

\item Additionally, if you are using existing assets (e.g., code, data, models) or curating/releasing new assets, \textbf{without compromising anonymity}...
\begin{enumerate}
  \item If your work uses existing assets, did you cite the creators?
    \answerYes{Yes, see Section \textit{Collection of Confounders} and \textit{Causal Inference of Emojis on Issues}}
  \item Did you mention the license of the assets?
    \answerNo{No}
  \item Did you include any new assets in the supplemental material or as a URL?
    \answerNo{No}
  \item Did you discuss whether and how consent was obtained from people whose data you're using/curating?
    \answerNo{No}
  \item Did you discuss whether the data you are using/curating contains personally identifiable information or offensive content?
    \answerNo{No}
\item If you are curating or releasing new datasets, did you discuss how you intend to make your datasets FAIR?
\answerNA{NA}
\item If you are curating or releasing new datasets, did you create a Datasheet for the Dataset? 
\answerNA{NA}
\end{enumerate}

\item Additionally, if you used crowdsourcing or conducted research with human subjects, \textbf{without compromising anonymity}...
\begin{enumerate}
  \item Did you include the full text of instructions given to participants and screenshots?
    \answerNA{NA}
  \item Did you describe any potential participant risks, with mentions of Institutional Review Board (IRB) approvals?
    \answerNA{NA}
  \item Did you include the estimated hourly wage paid to participants and the total amount spent on participant compensation?
    \answerNA{NA}
   \item Did you discuss how data is stored, shared, and deidentified?
   \answerNA{NA}
\end{enumerate}

\end{enumerate}

\appendix

\input{sections/9_appendix}

\end{document}

%% file: sections/1_introduction.tex
\section{Introduction}
\label{sec:intro}

Hybrid and remote working was considered a norm in the (post-)pandemic era. Yet, many people are concerned about the low efficiency in remote working, and some companies have launched the ``return-to-office'' policy to force employees to resume in-person working.\footnote{\url{https://www.bloomberg.com/news/articles/2022-11-10/musk-s-first-email-to-twitter-staff-ends-remote-work}, retrieved May 2023.} How to increase working efficiency in a virtual workspace has been a pressing problem. Some of the low efficiency can be attributed to the communication inconvenience in the virtual workspace. Despite the limited time on virtual meeting, text-based messaging is still a major format of communication in virtual workspace \cite{blanchard2021effects}. Missing in text-based communication are non-verbal cues such as facial expressions and body language, which convey subtle information about speakers' intent and sentiment, reduce misunderstanding, and are critical to successful communication. Emoji, as alternative non-verbal cues and prevalent in online communication, could be crucial to an effective discussion in the virtual workspace.

Indeed, beyond social network platforms, such as Twitter and Instagram, emojis are also gaining popularity on professional online platforms, such as GitHub.
As the largest host of source code in the world, GitHub offers distributed version control and source code management via the Git protocol. On GitHub, many distributed development activities are coordinated through \emph{issues},\footnote{\url{https://help.github.com/en/articles/about-issues}, retrieved May 2022.} which can be posted by any user to report software bugs, enhancement suggestions, or to solicit help. Conversations organized through issues and their comments can directly influence the quality of projects \cite{kavaler2017perceived}. Adequate and timely response to an issue is critical for the solution of the problem and the improvement of work outcomes. 

Emojis have been supported in GitHub issues as early as 2014 (see an example in Figure \ref{fig:issue_body_example}), and has become increasingly popular in recent years. In June 2017, 0.58\% of the issues contain emojis \cite{lu2018first}, while the ratio is 1.71\% in our collected issues in June 2021. 

Despite the increasing popularity, little is known whether using emojis benefits the developer community. That is, does using emojis improve the ``outcome'' of an issue, and does using more emojis bring more work outcomes for developers? 
%wei: what is work outcomes in this paper?
Research on emoji usage on social media platforms provides supportive evidence, citing emojis as ideal nonverbal cues to express sentiment, strengthen expression, adjust tone \citep{hu2017spice}, and engage the audience \citep{cramer2016emojifunction} in online communication, where facial expressions or body gestures are not available. 
However, in the more technical and professional conversations on GitHub, semantics and distribution of emojis differ from that used by the general population. Table \ref{tab:emoji_rank} shows the 20 most frequently used emojis in our collected GitHub issues (details in Section \textit{GitHub Issues and Problem Formulation}) and those released by the Unicode organization,\footnote{\url{https://home.unicode.org/emoji/emoji-frequency/}, retrieved in May 2022.} which ranks emojis based on their median frequency of use across multiple sources, representing the emoji usage in a more general context. Only five emojis (\MyEmoji{\partypopper}, \MyEmoji{\pray}, \MyEmoji{\smilingfacesmilingeyes}, \MyEmoji{\thumbsup}, and \MyEmoji{\grinningface}) overlap in the two lists. Several of the most popular emojis on GitHub have domain-specific meanings, such as \MyEmoji{\bugemoji} (bugs) and \MyEmoji{\rocket} (feature launching), while the popular emojis identified by the Unicode Consortium are mostly emotions and facial expressions. The gap in the emoji distribution is consistent with that reported in \citet{lu2022emojis}.

In this work, we take the initiative to quantify the effect of emojis in promoting participation in the developer community. We ask whether emojis attract more user participation to issues and eventually help resolve them. Moreover, since issues may serve different functionalities on GitHub, we also explore the heterogeneous effects of emojis across different issue content. We summarize our hypotheses below:

\begin{figure}[t]
\centering
    \includegraphics[width=\linewidth]{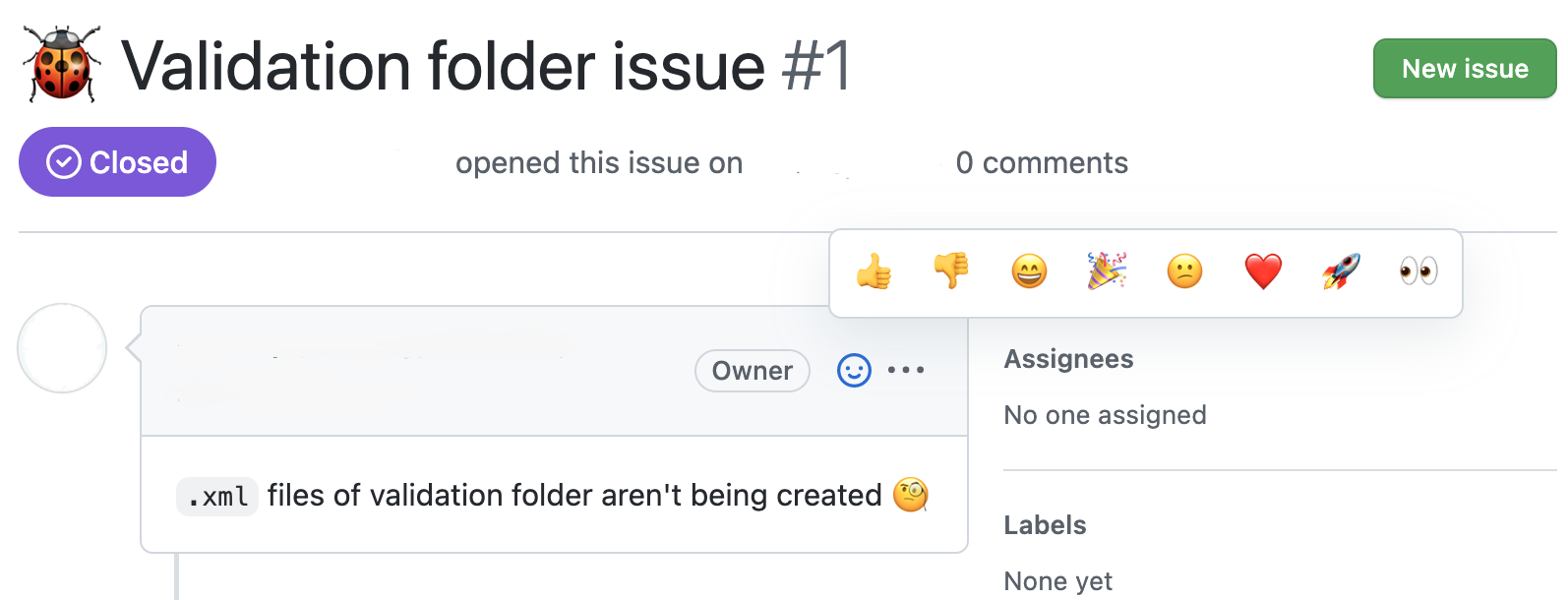}
    % \vspace{-0.5em}
    \caption{An example of emoji usage in GitHub Issues. The author puts \MyEmoji{\ladybugemoji} (lady beetle) in the issue title and \MyEmoji{\facewithmonocle} (face with monocle) in the issue body.}
    \label{fig:issue_body_example}
\end{figure}

\begin{description}
    \item[H1] Using emojis in an issue increases the participation of GitHub users.
    \item[H2] Issues with emojis are more likely to be resolved and resolved in a shorter time period.
    \item[H3] There exist heterogeneous effects by issue content.
\end{description}

To verify our hypotheses, we use causal inference techniques to estimate the causal effect of emoji usage on issues. We first collect a dataset of GitHub issues. Next, we construct the confounders that may influence both the application of emojis and the resolution / participation of the issue. We then apply propensity score matching to quantify the causal effect of emojis. Besides estimating the average treatment effect (ATE), we also perform a fine-grained analysis to quantify the heterogeneous effect by issue types. Finally, we conduct a case study to observe how emojis lead to causal effect.

\begin{table}[t]
    \centering
    \small
    \begin{tabular}{ l  l }
        \toprule
        Platform & 20 Most Frequently Used Emojis \\
        \midrule
        \multirow{2}{*}{GitHub Issues} &  \MyEmoji{\thumbsup} \MyEmoji{\stopsign} \MyEmoji{\smilingeyes} \MyEmoji{\rocket} \MyEmoji{\wavinghand} \MyEmoji{\partypopper} \MyEmoji{\whitecheckmark} \MyEmoji{\bugemoji} \MyEmoji{\warningemoji} \MyEmoji{\thinkingface}  \\
        & \MyEmoji{\sparkles} \MyEmoji{\slightsmilingface} \MyEmoji{\heart} \MyEmoji{\grinningface} \MyEmoji{\ladybugemoji} \MyEmoji{\smilingfacesmilingeyes} \MyEmoji{\moneybag} \MyEmoji{\pray} \MyEmoji{\triangular} \MyEmoji{\bigeyes}\\
        \midrule
        \multirow{2}{*}{Unicode Consortium} & \MyEmoji{\joy} \MyEmoji{\heart} \MyEmoji{\rollinglaughing} \MyEmoji{\thumbsup} \MyEmoji{\sob} \MyEmoji{\pray} \MyEmoji{\kiss} \MyEmoji{\smilingfacewithhearts} \MyEmoji{\hearteyes} \MyEmoji{\smilingfacesmilingeyes} \\
        & \MyEmoji{\partypopper} \MyEmoji{\beamingface} \MyEmoji{\twohearts} \MyEmoji{\pleadingface} \MyEmoji{\grinningface} \MyEmoji{\fire} \MyEmoji{\smilingface} \MyEmoji{\facepalming} \MyEmoji{\heartsuitemoji} \MyEmoji{\shrugging} \\
        \bottomrule
    \end{tabular}
    % \vspace{-0.5em}
    \caption{20 Most Frequently Used Emojis in GitHub Issues and reported by Unicode Consortium. We exclude emojis used as part of a template predefined by repository owner.}
    % \vspace{-1em}
    \label{tab:emoji_rank}
\end{table}

%% file: sections/2_related.tex
% \vspace{-0.5em}
\section{Related Work}
\label{sec:related}

Our work is mainly built on two streams of existing literature: the emoji functionalities and the GitHub issues.
\subsection{Emoji Functionalities}
The prevalence of emojis has led researchers to study their functionalities. Most previous research focuses on analyzing the emoji function from the semantic level to the audience level on social media \cite{hu2017spice, ai2017untangling, cramer2016emojifunction, zhou2022emoji, zhou2024emojis}, and only a few studies are conducted on professional platforms. In general, emojis are used to decorate texts \cite{miller2017understanding}, adjust tones, provide additional emotional information, and engage the audience \cite{cramer2016emojifunction}. These emoji-related research works mostly collect data from social media platforms, and researchers have identified the differences in emoji usage between communities and versions, such as apps, languages, cultures, genders, and platforms \cite{tauch2016emojirole, lu2016emojiusgae, chen2018gender, barbieri2016emojiusage, zhou2024adoption}. 

Several research studies notice the uniqueness of emojis on professional platforms, such as GitHub, and explore the intention of emojis. \citet{lu2018first} first studied the emoji distribution on GitHub and used manual annotation to infer the intention of emoji usage. This study shows that the fraction of GitHub issues with emojis increases sharply after March 2016. Recently, researchers have also applied machine learning models to automatically label the intention of using emojis in GitHub posts to help project maintainers monitor the quality of communication \cite{rong2022empirical}. Moreover, \citet{lu2023team} also discusses the emoji effects on team resilience based on GitHub data.
% \textcolor{red}{
% In addition, \cite{son2021more, borges2019beyond} explore the function of emoji reactions on pull requests and issues. They find that issues with giving emoji reactions (Figure \ref{fig:issue_body_example}) is statistically correlated with more discussions under pull requests and issues. 
% }
Researchers have also used emojis in GitHub issues for downstream tasks in software engineering, such as predicting developers' dropout \cite{lu2022emojis}, improving sentiment analysis \cite{chen2021emoji}. We extend this line of literature by quantifying the \emph{causal effect} of using emojis on the outcome of GitHub issues, and specifically the resolution and user participation of GitHub issues. For other causal inference work on the NLP application, none of them utilize this technique on GitHub platform and emojis \cite{feder2021causal, eckles2017bias}.

\subsection{User Engagement in GitHub Issues}
Many previous studies in the software engineering area focus on the sentiment analysis in social coding platforms. By exploring the relationship between sentiments on the social coding platform and developer productivity, researchers believe that understanding the relationship can help improve developer communication and production \cite{novielli2019sentiment, sanei2021impacts}. For issue discussions, it is widely recognized that the sentiments in issues can affect the productivity of developers \cite{mantyla2016mining}. Researchers analyze sentiments in issue discussions and find that the presence of emotions, especially positive emotions, is correlated with a shorter issue life cycle \cite{murgia2014developers, ortu2015bullies}, such as its resolution time. Researchers have also verified that sentiments in issues are correlated with their social engagement, such as follow-up discussion and response \cite{sanei2021impacts}. Taking into account emojis' functionality of expressing sentiments and adjusting tones \cite{hu2017spice}, we hypothesize that using emojis in issues can affect both the follow-up discussion and resolution of the issue.  

A common way to attract user participation in software development is by adding social signals, such as the Stack Overflow reputation score, GitHub badges, and GitHub followers \cite{trockman2018adding, tsay2014influence, merchant2019signals}.
These signals are used as indications of developers' expertise and commitment, which reduces users' assessment cost and promotes more user participation \cite{trockman2018adding}.
Literature has already verified emojis for their social signaling function. \citet{roberston2021black} has shown that skin-toned emoji can be regarded as a signal to provide users' identities and readers can perceive it. We hypothesize that emojis in GitHub issues are also social signals that help the audience perceive hidden information in GitHub issues, increasing the transparency of content to promote participation.

%% file: sections/3_data.tex
\section{GitHub Issues and Problem Formulation}
\label{sec:data}

\begin{table}[t]
    \centering
    \begin{tabular}{lrr}
        \toprule
         & Collected Issues & Issues w/ Emojis \\
        \midrule
        \# Issues & 203,098 & 14,686 \\
        \# Unique Users &  99,062 & 9,398\\
        \# Unique Repos &  90,115 & 9,185\\
        \bottomrule
    \end{tabular}
    \caption{Statistics of the datasets used for causal inference.}
    % \vspace{-1em}
    \label{tab:data_stats}
\end{table}

\subsection{Data Collection}
In our experiment dataset, we collect all GitHub issues in public repositories between June 1, 2020 and June 19, 2021 via a third-party project, GHTorrent.\footnote{\url{https://ghtorrent.org/}} GHTorrent monitors GitHub public event timeline. When someone creates, closes, or comments on an issue, GHTorrent retrieves the event content and dependencies, such as the related user and repository profiles. Only a small portion (1.6\%) of issues use emojis, but they represent a growing trend, so we are interested in estimating the effect of using emojis for these ``emoji issues'', also known as Average Treatment Effect on the Treated, or ATT. To make the dataset trackable, we sample issues with or without emojis using a ratio of 1:10 from the raw dataset. Such a ratio ensures that most treated issues find the untreated issues with similar propensity scores, while ensuring computational efficiency for further processing. 

In this work, we focus on the emojis used intentionally by the issue authors. But in practice, an issue may contain emojis because the repository sets up templates that the issue authors must follow.  The use of templates confounds the estimation of the causal effect of emojis on GitHub issues, but since the template format is determined by the repository owner, it is difficult to model ``template'' representations through the embedding method. Also, many issues are written by bots and do not reflect users' intention. So we remove all the issues with predefined templates or posted by bots. More details of the data preprocessing, detecting templated issues and bots are described in Appendix. After preprocessing and filtering, we construct a dataset of 203,098 issues where 14,686 issues contain one or more emojis. Table \ref{tab:data_stats} reports more detailed statistics of the used issues.

% \blue{we sample because (1) we care about average treatment effect on treated (ATT), (2) we want to make the data set trackable given there are way more non-emoji issues.}

\subsection{Outcome Variables}
\label{sec:outcome_variables}

Given the available data from GHTorrent and consistent with previous studies on issue discussions \cite{sanei2021impacts}, we use the comments below each issue, the closing status and time of an issue, to represent the participation and issue resolution, respectively. Specifically, we construct the following outcome variables:
\begin{description}
    \item[Developer Participation (H1)] We use whether getting comments, the number of comments, and the number of unique commenting users to measure the user participation level.
    \item[Issue Resolution (H2)] We calculate whether an issue is closed in 30, 60, 90, and 180 days, and the time before closing (for issues closed in 180 days after being posted) to measure the likelihood and speed of issue resolution. 
\end{description}

\begin{figure}[t]
    \centering
    \includegraphics[width=0.85\linewidth]{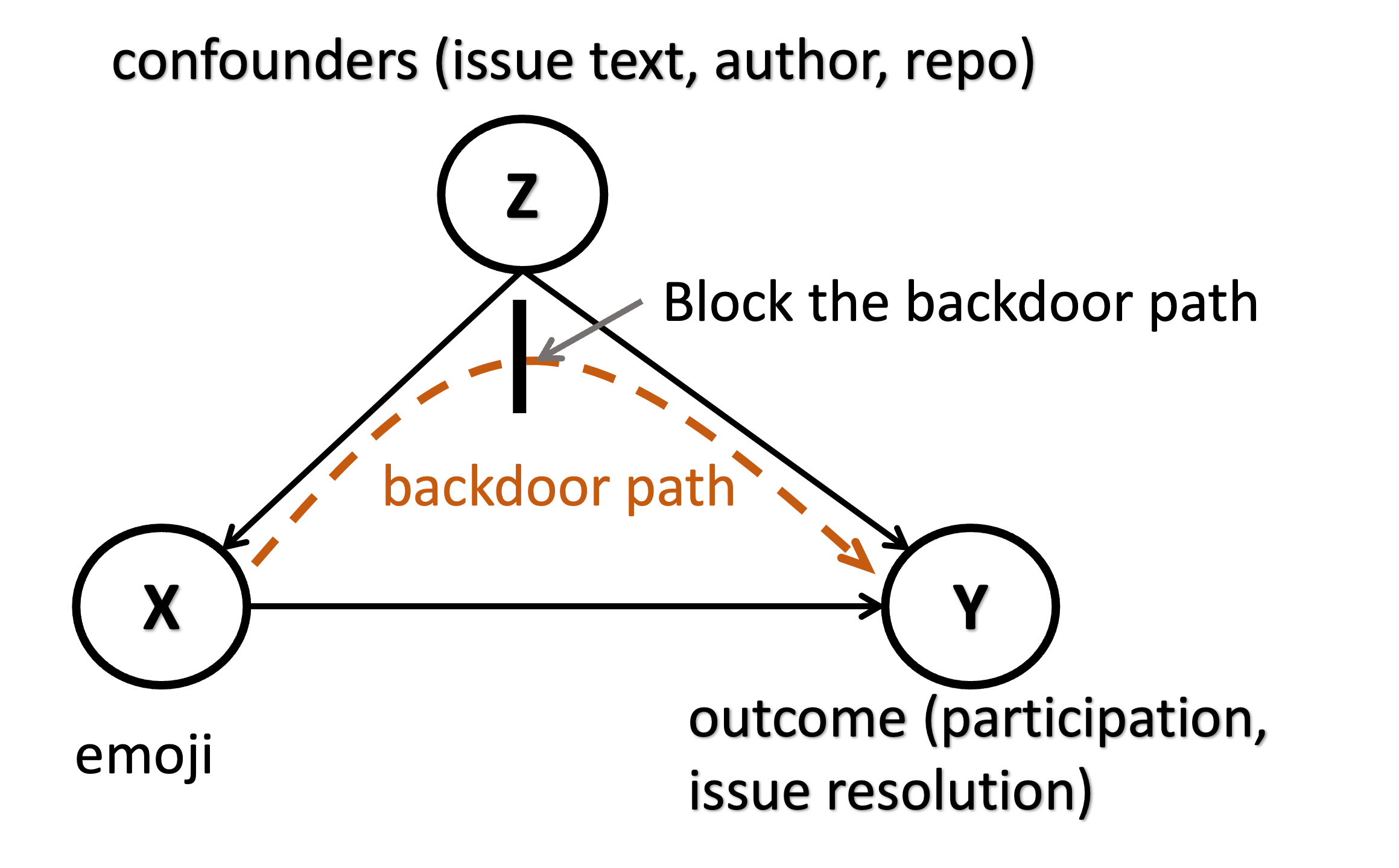}
    % \vspace{-0.5em}
    \caption{Causal graph for the backdoor adjustment}
    % \vspace{-1.0em}
    \label{fig:backdoor}
\end{figure}

Ideally, the observed difference of the outcome variables between the treatment group (issues with emojis) and the control group (issues without emojis) is the effect of using emojis. However, both the emoji usage and the outcome might be influenced by the same confounders, such as the issue contents and the authors. Therefore, instead of reporting the observed average difference between two groups, we use a principled causal inference technique to estimate the treatment effect of using emojis.
\subsection{Propensity Score Matching}
Formally, our objective for this project is to estimate the causal effect of emoji usage (binary cause $X$) on the issue resolution or user participation ($Y$) with multiple confounders ($Z$), such as the length of the issue, the topic of the issue, and the popularity of the repository. 
The relationship of the cause, outcome and confounder relationship are visualized in Figure \ref{fig:backdoor}. 
We use do-calculus language \cite{pearl1995causal} to quantify the causal impact of emojis. By mathematical formulation, our aim is to estimate $P(Y | do(X))$.

Since we cannot observe the counterfactual outcome of adding or deleting emojis for an issue, we can only estimate the causal effect by comparing observable issues with or without emojis. To quantitatively measure the effect of emoji usage, since the confounders can influence the outcome, we apply the backdoor adjustment \cite{pearl1995causal} to block every back-door path between $X$ and $Y$. Controlled confounders $Z$ should not be descendants of $X$. When $Z$ meets the backdoor criterion,
\begin{equation*}
    P(Y | do(X=x)) = \sum_z P(Y | X=x, Z=z)P(Z=z)
\end{equation*}
The average treatment effect (ATE) of $X$ (emojis) on $Y$ (that is, the closing time or the comment number) is
\begin{equation*}
    \begin{aligned}
     ATE & = \mathbb{E}[Y | do(X=1)] - \mathbb{E}[Y | do(X=0)] \\
        % & = \sum_Z (\mathbb{E}[Y | X=1, Z] - \mathbb{E}[Y | X=0, Z]) P(Z) \\
        & = \mathbb{E}_Z[\mathbb{E}[Y | X=1, Z] - \mathbb{E}[Y | X=0, Z]]
    \end{aligned}
\end{equation*}
Although we can follow the equations to calculate the treatment effect, in our scenario, $Z$ is a high-dimensional vector that represents all confounders. It is difficult to find an issue containing emojis and an issue not containing emojis, but with the same values of $Z$. \cite{rosenbaum1983central} has proposed an alternative way to calculate causal effect by matching not on $Z$, but on the propensity score $R = P(X=1 | Z)$, which is called propensity score matching (PSM). The equations of calculating ATE become
\begin{equation*}
\label{PSM}
    ATE = \mathbb{E}_R[\mathbb{E}[Y | X=1, R] - \mathbb{E}[Y | X=0, R]]
\end{equation*} 

Therefore, we need to match issues containing emojis with issues having similar $R$ but not containing emojis. To estimate the propensity score $R = P(X=1 | Z)$ for each issue, we need to observe all possible confounders $Z$ that can influence both the usage of emojis and the outcome of the issues. 

%% file: sections/4_confounders.tex
% \vspace{-0.8em}
\section{Collection of Confounders}
To control for the confounder data of issues, we construct the following covariates based on available data from three sources: the text of the issue, the author of the issue, and the repository profile.

\begin{table}[t]
    \centering
    \footnotesize
    \begin{tabular}{ l  m{18.5em} }
        \toprule
        Category & Confounder Variables \\
        \midrule
        Issue Text & created week, \# of characters in the title, \# of tokens in the title, \# of characters in the body, \# of tokens in the body, politeness score, topic distribution, issue type labels \\
        \midrule
        Issue Author & author account age (days), \# author followers, \# author following, \# public repos of the author \\
        \midrule
        Repository & repo age (days), \# repo stars, \# repo forks, \# open issues in the repo, \# repo watch \\
        \bottomrule
    \end{tabular}
    % \vspace{-0.5em}
    \caption{Confounders collected for the causal effect of emoji usage.}
    \label{tab:confounder_stats}
\end{table}

\subsection{Issue Text Confounders}
\label{sec:issue_confounders}
For issue covariates, we collect its creation time, length of its title and body (measured as the number of characters and tokens). In addition, to control for the semantics of the issues, we measure the issue type, issue topic, and issue politeness as proxies for the semantics of the issue.

\begin{figure}[t]
    \centering
    \includegraphics[width=\linewidth]{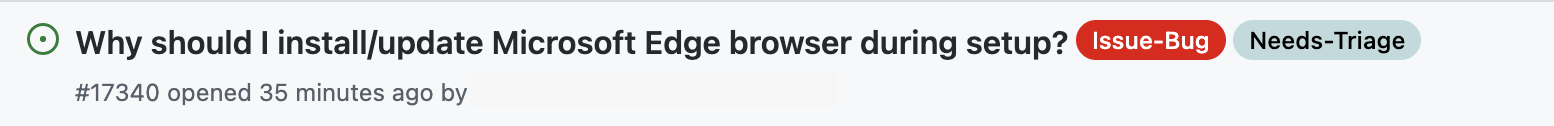}
    \caption{The example of an issue with self-annotated labels}
    % \vspace{-1.5em}
    \label{fig:issue_types}
\end{figure}

\textbf{Issue Types} Authors post issues for various reasons, such as reporting bugs, seeking help, or requesting new features, and they sometimes label issue types explicitly with self-annotated tags. Figure~\ref{fig:issue_types} shows an example where the author labeled the issue with ``Issue-Bug'' and ``Needs-Triage.'' 

According to the data from GHTorrent, the 3 mostly frequently used and content-related labels in GitHub issues are \emph{bug} (reporting bugs), \emph{question} (asking for help), and \emph{feature} (requesting new features). 
Using the above three self-annotated labels as ground-truth, we train three classifiers to automatically label whether the issue belongs to bug, feature, or question type. Details of the classifiers are given in Appendix B. 

\textbf{Topics in Issues} 
Besides issue types, issues also contain multiple topics, such as game, programming language, or operating system. To capture the latent topics, we use the unsupervised topic modeling, the Latent Dirichlet Allocation (LDA) model, to generate the representation of the topic \cite{blei2003latent}. The reason we do not directly use contextualized document embeddings, such as BERT embeddings \cite{devlin2018bert}, as the confounder is that there is no clear interpretation of each dimension of document embedding, and it is hard for us to check whether this confounder is balanced after matching. Therefore, we chose LDA topic embeddings to represent the confounder of issue content.
After enumerating the topic number as 10, 20, 30, and 40 to train multiple LDA models, we observe that when setting the topic number as 30, the keywords (words with the highest weights) of each topic are the most interpretable, and the number of overlap keywords between topics is the smallest.
Thus, for each issue, we obtain a 30-dimensional topic distribution vector, where each dimension is the distribution of the topic in the issue.

\textbf{Politeness in Issues} Since emojis have the functionality to signal politeness \cite{escouflaire2021signaling} and politeness of issues may encourage issue resolution and user participation, we measure the issue politeness score as a confounder. \citet{danescu2013computational} has proposed a politeness classifier with domain-independent lexical and syntactic features, which is trained on Wikipedia and Stack Exchange data and achieves 78.19\% on their Stack Exchange dataset. For each issue, we apply the pretrained classifier to calculate the politeness score as the issue's politeness.

\subsection{Issue Author and Repository Confounders}
Just as we need to control for the demographics for human subject, we should also control the ``demographics'' of an issue. Therefore, we extract author and repository information of an issue, such as their popularity and primary programming language, which may confound the treatment (using emoji) and outcome. For authors, we collect the number of followers and followings, the number of public repositories, and the account age. For repositories, we collect the number of stars, the number of forks, the number of open issues, the programming language used, and the repository age. Table \ref{tab:confounder_stats} summarizes our collected covariates.

%% file: sections/5_method.tex
\section{Causal Inference of Emojis on Issues}
\label{sec:causal_inference}
\subsection{Propensity Score Estimation}
In our scenario, the propensity score $R$ is the probability of issues using emojis given confounders ($Z$). Following the econometrics literature, we choose a parametric model, namely logistic regression (LR), to estimate the propensity score \cite[Chapter~3]{angrist2008Mostly}.
Our LR will predict whether the issue contains emojis given confounder variables $Z$, and the probability of inference on each issue is the estimated propensity score. Since most of the issues in our dataset do not contain emojis, the model can easily achieve high accuracy by only majority class selection. To address the class imbalance, we first perform undersampling to ensure the same number of issues with and without emojis in the training dataset.
After training, we apply the model on the original unbalanced dataset and find matched untreated issues for each treated issue.

Evaluating the propensity score estimation is tricky as we cannot use the out-of-sample accuracy. An issue is either treated or untreated in the dataset, and we do not know the ground truth of its probability of being treated. Instead, we rely on the balance check (detailed in Section \textit{Nearest Neighbor Matching}) to examine whether the matched samples (of both treated and untreated) are similar or comparable other than their treatment status. Speciﬁcally, we want to check whether the covariate distribution is balanced between the matched samples. 
\subsection{Nearest Neighbor Matching}
\label{sec:nn_matching}

Since the assessment of the propensity score estimation is closely related to our chosen matching method, we first introduce the nearest neighbor matching. Here, we choose the most common implementation for the propensity score matching, pair matching \cite{austin2021applying} (one-to-one matching). We pair each treated issue with an untreated issue and then average the difference of the matched pairs as the average treatment effect on treated (ATE). Literature suggests that matching with replacement may lead to the condition that an untreated issue is used in multiple pairs, which may cause the variance estimation of the treatment effect to be more complex \cite{hill2006interval}, we implement matching without replacement in this paper. Once an untreated issue is selected to match a treated issue, it will no longer be used to pair with other treated issues.

We implement pair matching using greedy nearest-neighbor matching (NNM) instead of optimal matching. Optimal matching finds pair matchings that minimize the sum of the score difference within each pair \cite{austin2021applying}. Although it can find the global optimal matching, it is computationally infeasible for our scale of data, and previous study has shown that optimal matching does not behave much better in forming balanced pairs \cite{gu1993comparison}. For greedy matching, we randomly select a treated issue and match it with an untreated one with the closest propensity score, and repeat the process for other treated issues.

We further ensure the quality of matching by applying a caliper restriction. That is, a matched pair is acceptable only if the within-pair difference of the propensity scores is less than the predefined caliper width. We discard treated issues that cannot be matched to untreated issues of similar propensity score. We set the caliper width to 0.2 of the standard deviation (SD) of the propensity score, which has been shown to work well in multiple settings \cite{austin2011optimal}. We also observe that the proportion of emoji usage for each issue type is not similar, so to keep a balanced distribution of issue types, we enforce the exact matching on issue types. That is, issues will only be matched to other issues of the same type. We will discuss issue types in more details in Section \textit{Heterogeneous Effects by Issue Types}.
%\blue{wei: is there other motivation than ``the proportion is not similar''? it doesn't sound very convincing.}

To assess the matching of the propensity score, we expect the distribution of the confounders in the treatment and control groups to be balanced. Standardized mean difference (SMD) is a commonly used measurement to examine the balance of the covariate distribution between groups \cite{zhang2019balance}. The guidelines indicate that 0.1 or 0.25 represent reasonable cutoffs and a higher value of SMD indicates that the covariate distribution in one group is too different from one another for reliable comparison \cite{stuart2013prognostic}. We visualize the SMD values of each covariate before and after matching in Figure \ref{fig:smd_value} in Appendix. The nearest neighbor matching greatly reduces the distribution difference between the confounders of issues with emojis and without emojis. The SMD values of confounders are all below the 0.1 threshold, showing the similar covariate distribution between groups. More details of SMD check and additional refinement are discussed in Appendix.

%% file: sections/6_result.tex
\section{Results and Discussion}
\label{sec:results}
\subsection{Average Treatment Effect}
\label{sec:ate}
By averaging the pairwise difference between treated and untreated issues, we estimate the unbiased treatment effect of emoji usage. We calculate the ATE of the outcome variables mentioned in Section \textit{Outcome Variables} and show the results in Table \ref{tab:outcome_result}. For the issue closing time variable, since GitHub repositories may automatically close some issues without any discussion for a long time and our dataset covers the issues over a one-year period, we cannot tell whether issues with long closing time are resolved. Therefore, we only consider the issue closing time less than 180 days, covering 98.6\% of all closed issues. Besides the ATE, we also report the observed average difference between two groups before matching (Obs. $\Delta$) as well as the average value in treatment groups (Avg.). Since we are measuring the average pairwise difference, we apply the paired $z$-test \cite{derrick2017test} and the differences are all statistically significant at the 1\% level.

\begin{table}[t]
\centering
\small
\begin{tabular}{lrrr}
\toprule
Outcome Variable                   & ATE & Obs. $\Delta$ & Avg. \\ \midrule
\textit{Developer Participation} & & & \\[0.3em]
getting comments                    & 0.041**    & 0.112                    &  0.476 \\
                                         \# of comments                        & 0.118**    & 0.346                    &  1.236 \\
                                         \# of comment users              & 0.113**    & 0.257                    &  0.811 \\
                                          \midrule
\textit{Issue Resolution}                   & & & \\[0.3em] 
\% closed in 180 days & 0.033**    &  0.056                   &  0.391 \\
                                         \% closed in 90 days  & 0.033**
    &  0.052                   &    0.378 \\
                                         \% closed in 60 days  & 0.035**   &  0.053                   &  0.370 \\
                                         \% closed in 30 days  & 0.038**
    &   0.051                  &   0.346  \\
                                         issue closing time (day)          & -1.751**    & -0.822                    &   11.570  \\ \bottomrule
\end{tabular}
% \vspace{-0.5em}
\caption{The treatment effect of using emojis in GitHub issues. Obs. $\Delta$: The observed average difference between the control group and treatment group without matching on propensity scores. Avg: the average value in the treatment group. Signiﬁcance level: ** $p<0.01$, * $p<0.05$, paired $z$-test.}
% \vspace{-1.5em}
\label{tab:outcome_result}
\end{table}

\paragraph{Developer Participation} We first examine whether emojis bring more discussion to an issue, which can be measured by the likelihood of getting comments or the number of comments. As shown in Table \ref{tab:outcome_result}, issues with emojis are 4\% more likely to receive comments. On average, issues with emojis get 0.118 more comments than those without emojis. We further check the number of users participating in the conversation and find that using emojis attracts 0.113 more users to participate in issue discussion. All results confirm our hypothesis H1 that using emojis in issues attracts more participation in issue discussions. 

\paragraph{Issue Resolution}
People do not simply want their issues to be watched; they want their issues to be resolved -- bugs ﬁxed, features added, and questions answered. Beyond the increased participation, does emoji usage actually help to resolve the issues?
By looking at the closing status of the issues at different time intervals, we find that issues with emojis are more likely to be closed in all time periods, and the effect is larger for the shorter time periods (ATE increases from 3.3\% to 3.8\% when interval decreases from 180 to 30 days). For issues that are closed in 180 days, we find that using emojis speeds up their resolution by an average of 1.751 days. Therefore, we may infer that the attention and participation that emojis attract are not from mere bystanders. Shorter closing time and larger proportion of closed issues with emoji use confirm our hypothesis H2.

\subsection{Sensitivity Analysis}
\label{sec:sensitivity}
A major threat to validity for propensity score matching is that there might be unobserved confounders influencing the estimated causal effect of emojis. We perform a simultaneous sensitivity analysis to examine the extent to which our conclusion is robust to unobserved confounders \cite{gastwirth1998dual}. The intuition is that if an unobserved confounder invalidates our conclusion, it should be strongly correlated with the treatment and/or the outcome. We may use the observed confounders as reference points for the strength of such correlation, and ask the question: if an unobserved confounder has the same strength of correlation as that of the observed confounders, what would the significance of the treatment effect become, that is, how large would the $p$-value become? If the new $p$-value is still small, it means that the treatment effect is still significant even with the presence of such an unobserved confounder.
Formally, we denote $\mathcal{T}$ as the upper bound of the odds ratio between the unobserved confounder and the treatment, and $\Delta$ as the upper bound of the odds ratio between the unobserved confounder and the outcome. The goal of the simultaneous sensitivity analysis is to see whether an unobserved confounder with a combination of $\mathcal{T}$ and $\Delta$ would make the estimated causal effect insignificant ($p \geq 0.05$) \cite{liu2013introduction, gastwirth1998dual}. Given the values of $\mathcal{T}$ and $\Delta$, we can calculate the upper bound $p$-value as follows:
\begin{align*} 
& p(\theta) =  \frac{\Delta}{1 + \Delta} \quad p(\pi) = \frac{\mathcal{T}}{1 + \mathcal{T}} \\
& p^+ = p(\pi) \times p(\theta) + (1-p(\pi)) \times (1-p(\theta)) \\
& \textit{upper bound P-value} = \sum_a^T \binom{T}{a} (p^+)^a (1-p^+)^{T-a}
\end{align*}
where $T$ is the total number of discordant pairs in which the outcomes differ within the pair and $a$ is the number of pairs in which the issue with emojis has an outcome and the issue without emojis does not \cite{liu2013introduction}. In our experiments, the values of $T$ and $a$ are 7,905 and 3,805 respectively. 

We err on the conservative side in fixing the values of $\mathcal{T}$ and $\Delta$ as the largest odds ratio between observed confounders and the treatment variable and the outcome variable, respectively \cite{liu2013introduction}. For binary observed confounders, we use their odds ratios with the treatment and outcome variables. For continuous observed confounders, we fit logistic regression models between the confounders and treatment or outcome, and use the learned model parameters to estimate the odds ratio \cite{bland2000odds}. 
% \textcolor{red}{The confounder with the largest odds ratio to the treatment is the label of bug issues. For outcome variables, the strongest confounder to the outcome ``getting comments" is the politeness and for other variables, the confounder with the largest strength is Topic 28, predicted by LDA model, about the website error.}
We report the values of $\mathcal{T}$ and $\Delta$ and the upper bound $p$-values for five binary outcome variables in Table \ref{tab:simultaneous_result}. 
We observe that all the upper bound $p$-values are less than 0.05. The results can be interpreted as that the estimated treatment effect of emojis is still significant, even if there exists an unobserved confounder which has the same strength of association with the binary outcome and the treatment as the strongest ones of all observed confounders.

Besides the sensitivity analysis on weakening the unconfoundedness assumption, we also verify the robustness of our findings by repeating the analysis on alternative specifications \citep{athey2017state}, such as using dataset from a different time span, choosing a different propensity score estimator, or changing the number of topics in the LDA topic model. We repeat the analysis on the GitHub issues created between July 2018 and June 2019, which is before the pandemic shock, and find that the treatment effects of using emojis have the same direction and similar significance levels to our main result. More details are shown in Appendix.
%athey_and_imbens_2017_state = The State of Applied Econometrics: Causality and Policy Evaluation
%to reproduce the causal inference. Details of another dataset and results are shown in  The treatment effect of emojis on issues from another-year duration shows the same pattern in significance and directions as the results in Table \ref{tab:outcome_result}, which verifies the robustness of our findings and avoids the effect of other external confounders such as COVID-19 pandemic. 
We also apply the Gradient Boosted Regression Tree method (rather than logistic regression) to re-estimate the propensity score, or change the topic number to 15 or 45 to retrain the LDA model and repeat the analysis. Both show similar patterns in the significance and directions of the estimated treatment effect (details in Appendix).
 %We reproduced results in the reproduction suggest that the estimated effect of emoji usage is robust to the propensity score estimation method or the topic number of LDA model.}

\begin{table}[t]
\centering
\small
\begin{tabular}{lrrrr}
\toprule
Binary Outcome Variable                   & $\Delta$ & $\mathcal{T}$ &  $p$-value \\ \midrule
\textit{Developer Participation} & & &  \\[0.3em]
getting comments                    & 4.290    & 1.163                   &  0.015   \\
                                          \midrule
\textit{Issue Resolution}                   & & &  \\[0.3em] 
\% closed in 180 days & 4.280    &  1.163                  &  0.015   \\
                                         \% closed in 90 days  & 4.188
    &  1.163                   &    0.013  \\
                                         \% closed in 60 days  & 4.542   &  1.163                   &  0.020     \\
                                         \% closed in 30 days  & 4.816
    &   1.163                  &   0.026   \\
\bottomrule
\end{tabular}
% \vspace{-0.5em}
\caption{Simultaneous sensitivity results for binary outcome variables. $\Delta$ denotes the largest odds ratio between the confounder and the outcome. $\mathcal{T}$ denotes the largest odds ratio between the confounder and the treatment. $p$-value is the calculated upper bound $p$-value.}
% \vspace{-1.5em}
\label{tab:simultaneous_result}
\end{table}

%% file: sections/7_implication.tex
% \vspace{-0.5em}
\subsection{Heterogeneous Effects by Issue Types}
\label{sec:fine-grained}
The results in Section \textit{Average Treatment Effect} show that using emoji can promote developer participation and speed up issue resolution. However, authors may post different types of issues, choose among thousands of emojis, and expect different kinds of responses. We wonder whether the emoji effect is consistent across issue types. In this section, we re-estimate the treatment effect of different issue types and explore the heterogeneous treatment effect (HTE) of emoji usage. We will follow the definition in Section \textit{Issue Text Confounders}, and classify issues with three types: bug, question, and feature. 

\begin{table*}[ht]
\centering
\small
% \begin{tabular}{l @{\extracolsep{1.3em}} c @{\extracolsep{1.3em}} c @{\extracolsep{1.6em}}c @{\extracolsep{1.3em}}c @{\extracolsep{1.3em}} c @{\extracolsep{1.3em}}c @{\extracolsep{1.3em}}c}
\begin{tabular}{l @{\extracolsep{1.3em}} c @{\extracolsep{1.3em}} c @{\extracolsep{1.6em}}c @{\extracolsep{1.3em}}c @{\extracolsep{1.3em}} c @{\extracolsep{1.3em}}c @{\extracolsep{1.3em}}c}
\toprule
& & \multicolumn{3}{c }{Heterogeneous Treatment Effect (HTE)} & \multicolumn{3}{c}{HTE on type-associated emojis} \\ 
\cmidrule{3-5}\cmidrule{6-8}
Outcome Variable    & ATE                               & Bug       & Question   & Feature & Bug  & Question & Feature \\
\cmidrule{1-1} \cmidrule{2-2}\cmidrule{3-5}\cmidrule{6-8}                   
getting comments  &     0.041**             & 0.052**  & -0.019          & 0.037** & 0.060**                & 0.011                      &  0.030**         \\
\# of comments  &   0.118**                   & -0.013             & -0.128* & 0.225** & -0.224**                  & -0.080                     & 0.301**        \\
\# of comment users   &   0.113**          & 0.042             & -0.059* & 0.208**  & -0.013                & 0.010                      & 0.264**         \\ 
%\cmidrule{1-1}\cmidrule{2-4}\cmidrule{5-7}
[1em]
\% closed in 180 days & 0.033**  & 0.031**   & 0.024*            & 0.007 & 0.053**                 & 0.036**                      & 0.005          \\
\% closed in 90 days & 0.033**   & 0.033**   & 0.027*            & 0.007  & 0.053**                 & 0.034**                      &  0.008         \\
\% closed in 60 days & 0.035**   & 0.036**  & 0.030*            & 0.009 & 0.054**                 & 0.036**                      &   0.014        \\
\% closed in 30 days & 0.038**   & 0.044**   & 0.026**            & 0.009 & 0.060**                 & 0.039**                       &  0.008        \\
issue closing time (day) &  -1.751**        & -2.673** & -0.378           & -1.313 & -2.688**                &  -0.470                     & -0.998          \\
[1em] 
\# Issues in Treatment Group &    14,686          & 5,201  & 4,227     & 6,441 & 1,784                    & 2,230                       & 3,445 \\
\bottomrule
\end{tabular}
% \vspace{-0.5em}
\caption{The treatment effect of using all emojis and positive PMI emojis in GitHub issues per issue types. The ATE column is the same with Table~\ref{tab:outcome_result} and is included here for reference. Signiﬁcance level: ** $p<0.01$, * $p<0.05$, paired $z$-test.}
% \vspace{-1em}
\label{tab:type_outcome}
\end{table*}

\sloppy Since we only calculate the treatment effect of a subset of the treatment group, the definition of propensity score also changes from $p(\text{issue contains emojis} | \text{confounders})$ to $p(\text{issues contains emojis} | \text{confounders}, \text{issue type})$. We retrain the logistic regression model to estimate the new propensity score within each issue type and perform the nearest neighbor matching again. With the caliper restriction, we are able to ensure that the SMD value for each confounder variable is less than 0.25. We then recalculate the average pairwise difference as the HTE and present the results in Table \ref{tab:type_outcome} (the left half).

The HTE shows an interesting pattern that aligns with the types of issues. For bug and question issues, the issue authors hope to get them fixed or answered in a timely manner. Indeed, we observe significant treatment effects on issue resolution (the closing status and time). On the contrary, the effect on participation is not consistent with the ATE. Using emojis in question issues even reduces the number of comments and comment users. One possible reason is that bug and question issues with emojis are closed in a shorter time, participants may have less chance to comment, which is reflected as a negative treatment effect. On the other hand, using emojis in feature request issues does not have the significant effect on its closing status, but significantly increases the comments it receives. Such effects may also be desired, as the authors of the feature request may not expect to close issues as soon as possible, but do hope to attract more participation to move the ideas forward.

To summarize the HTE for different issue types, we find that using emojis benefits issue authors in achieving their \emph{desired outcome}. 

\subsection{Emojis as Social Signals}

Indeed, such heterogeneity in issue types sheds light on the possible explanations why emojis promote developer participation and issue resolution. As discussed in Section \textit{Related Work}, social signals \cite{spence2002signaling} have been shown to be useful in attracting user participation. For example, Stack Overflow reputation scores, GitHub badges, and GitHub followers can signal a developer's expertise and commitment, which reduces users' assessment cost \cite{trockman2018adding, tsay2014influence, merchant2019signals}. Similarly emojis can also be regarded as observable signals to reduce the information asymmetry in online communication.
We hypothesize that emojis on GitHub, as signals, can signal the overall topic and the author's attitude. By using emojis, the readers can better perceive the author's attitude and intent, which can reduce information asymmetry between authors and readers, and attract readers to participate in the issue resolution.

If such a hypothesis is correct, we should expect issue authors to purposefully select the emojis that signal their intent, and issues with the ``right'' emoji signals are more likely to have better outcome. Inspired by the observation of different emoji distribution between Unicode and GitHub in Table \ref{tab:emoji_rank}, we may expect different issue types are associated with different emojis. Using the emojis that are associated with specific types of issues serves as the social signals that promote the desired outcome. 

To this end, we first calculate the association between the choice of emojis and the types of issues, using the point-wise mutual information (PMI) between emojis and issue types \cite{church1990word}. For each emoji $e$, the PMI value for the type of issue $i$ is $\text{PMI}(e, i) = \log \frac{p(e, i)}{p(e)p(i)}$.

A positive PMI indicates the positive association between the emoji and the issue type. The top 10 most frequently used emojis with positive PMI for each issue type are shown in Table \ref{tab:emoji_pmi}. We find that bug issues are positively associated with emojis with domain specific meanings such as \MyEmoji{\bugemoji}, \MyEmoji{\ladybugemoji} and \MyEmoji{\checkmarkemoji}. Positive PMI emojis for question issues are mostly emotions and facial expressions. For feature issues, nearly half of the emojis are domain-specific (\MyEmoji{\rocket}, \MyEmoji{\moneybag} and \MyEmoji{\triangular}) and half express emotions.

\begin{table}[t]
    \centering
    \small
    \begin{tabular}{ l l }
        \toprule
        Issue Type & 10 Most Frequently Used Emojis \\
        \midrule
        Bug &  \MyEmoji{\stopsign} \MyEmoji{\bugemoji} \MyEmoji{\warningemoji} \MyEmoji{\ladybugemoji} \MyEmoji{\checkmarkemoji} \MyEmoji{\crossmarkemoji} \MyEmoji{\clipboard} \MyEmoji{\speechballon} \MyEmoji{\openhands} \MyEmoji{\robot} \\

        Question & \MyEmoji{\thumbsup} \MyEmoji{\smilingeyes} \MyEmoji{\wavinghand} \MyEmoji{\thinkingface} \MyEmoji{\slightsmilingface} \MyEmoji{\heart} \MyEmoji{\grinningface} \MyEmoji{\smilingfacesmilingeyes} \MyEmoji{\pray} \MyEmoji{\bigeyes} \\
        
        Feature & \MyEmoji{\thumbsup} \MyEmoji{\smilingeyes} \MyEmoji{\rocket} \MyEmoji{\partypopper} \MyEmoji{\whitecheckmark} \MyEmoji{\sparkles} \MyEmoji{\heart} \MyEmoji{\smilingfacesmilingeyes} \MyEmoji{\moneybag} \MyEmoji{\triangular} \\
        \bottomrule
    \end{tabular}
    % \vspace{-0.5em}
    \caption{10 Most frequently used emojis with positive PMI for each issue type.}
    % \vspace{-1.5em}
    \label{tab:emoji_pmi}
\end{table}

Now that we have identified emojis associated with each issue type, we repeat the estimation of the HTE but restrict to issues with the type-associated emojis. 
% We only keep the issues containing emojis with positive PMI value, repeat the NNM (maintain SMD values less than 0.25) process, and obtain the ATE value of the outcome variables. 
We report the HTE of using type-associated emojis on the right side of Table \ref{tab:type_outcome}. Compared to the HTE estimated on all issues, the HTE on type-associated emojis is in the same direction but at larger scales. %original causal effects in each issue type group (the left side results in Table \ref{tab:type_outcome}), positive PMI emojis can make the original significant ATE more remarkable. 
For bug and question issues, type-associated emojis can further reduce the resolution time. For example, the increase in the proportion of issues closed in 180 days increases from 0.031 to 0.053 and from 0.024 to 0.036 for bug issues and question issues, respectively. Similarly, for feature issues, type-associated emojis promote more active participation, as the treatment effect on the number of comments increases from 0.225 to 0.301. These findings show that the treatment effect is larger when authors choose emojis that signals the type of the issues.

The analysis of HTE on type-associated emojis presents only preliminary findings about social signals as a possible mechanism that explains causal effect of using emojis. Future efforts are needed to fully examine emojis as social signals.  Other possible explanations may also hold. 
For example, one known functionality of emojis is adjusting the tones and sentiments \cite{hu2017spice, cramer2016emojifunction}, and there also exists the correlation between sentiments and issue outcomes \cite{sanei2021impacts, murgia2014developers, ortu2015bullies}. It is possible that emojis first adjust the affective states (sentiments and tones) in issues, and the adjusted affective states bring more attention, which leads to more participation.

\subsection{A Case Study}

\begin{table}[!t]
    \begin{subtable}{1\linewidth}
    \footnotesize
    \centering
        \begin{tabular}{ m{4.5em} m{19.5em} }
            \toprule
            Issue Title & Hello \\
            \midrule
            Issue Body & @[user] I currently can't access discord because I don't have a phone number to verify that I'm a real human. \MyEmoji{\pensiveface}\\ \midrule
            Comments & Use that free SMS service \\
            \bottomrule
        \end{tabular}
        \caption{A question issue and the developers' comments. This issue is closed in 1 day and contains 1 comment below.}
    \end{subtable}
    % \vspace{-0.3em}
    
    \begin{subtable}{1\linewidth}
    \footnotesize
    \centering
        \begin{tabular}{ m{4.5em}  m{19.5em} }
            \toprule
            Issue Title & Add Portainer software docs \\
            \midrule
            Issue Body & [pull request url] Need to add Portainer docs before 6.34 release \MyEmoji{\wink}  \\ \midrule
            Comments & Ok linking the page as a basic help looks good for me. :-) \newline [url] well usually we don't add fill software docs. \newline
            More basic thinks like how to access. But we could link this guide or the official docs\\
            \bottomrule
        \end{tabular}
        \caption{A feature issue and the developers' comments. This issue is closed in 12 days and contains 3 comments below.}
    \end{subtable}
    % \vspace{-0.3em}
    
    \begin{subtable}{1\linewidth}
    \footnotesize
    \centering
        \begin{tabular}{ m{4.5em} m{19.5em} }
            \toprule
            Issue Title & Docs readme typo \\
            \midrule
            Issue Body &  \MyEmoji{\crossmarkemoji} `index.js' \newline  \MyEmoji{\checkmarkemoji} `index.html' \\ \midrule
            Comments & Thanks for pointing that out, [at mention] .\\
            \bottomrule
        \end{tabular}
        \caption{A bug issue and the developers' comments. This issue is closed in 0 days and contains 1 comment below.}
        % \vspace{-0.3em}
    \end{subtable}
    \caption{Issues examples with emojis. Comment number and closing days are listed below each sub-table.}
    % \vspace{-1em}
    \label{tab:case_study}
\end{table}

To better understand how emojis apply functions, we conduct a case study on 3 selected issues, where authors use emojis. Table \ref{tab:case_study} lists the titles of the issue, the bodies of the issue, and the attached comments. All the issues are closed and are automatically labeled as the question, feature and bug issue respectively. 
% In the first question issue, the authors use \MyEmoji{\worriedface} (worried face) in the title of the issue. From the sentiment adjustment aspect, \MyEmoji{\worriedface} shows the author's disappointment and unhappiness emotion. 
% If removing this emoji, the overall sentiment of the issues tends to be neutral. The emoji also indicates the author's disappointed attitude towards the missing license, and the hidden anxious attitude may push the repository owners to answer the question. From the comment below this issue, the developer obtains the authors' attitude, so emoji promotes the issue resolution. 

In the first question issue, the authors use \MyEmoji{\pensiveface} (pensive face) in the issue body. From the sentiment adjustment aspect, \MyEmoji{\pensiveface} shows the author's disappointment and unhappiness emotion. 
If removing this emoji, the overall sentiment of the issues tends to be neutral. The emoji also indicates the author's frustration towards the inaccessibility to discord, and the hidden anxious attitude may nudge the repository owners to answer the question. From the comment below this issue, the developer notices the authors' attitude and offers a solution, so emoji promotes the issue resolution.

For the second feature issue, the author proposes a new pull request to add a new feature and applies \MyEmoji{\wink} to the issue body. This emoji suggests that the author is satisfied with the new feature and the emoji makes the issue sentiments to be positive. Again, from the three comments below, the other developers receive the information in the signal and appreciate the author's contribution. 

For the third bug issue, the issue is about to fix a typo, and the author only uses two emojis to signal where to change. Although unlike two issues above, the semantics of this bug issue can also be expressed by words, the emojis make the issue more concise and formalized, signaling that the author would like to save the other developers' time, which shows consideration for readers' comfort. This issue is closed immediately, and the typo is also fixed.

% \vspace{-0.2em}
\section{Limitations and Implications}
\label{sec:limitation}
There are several limitations for our work. First, a major threat to validity for using propensity score is the unobserved confounders. Due to the limitations of the dataset, we cannot fully capture all covariates of an author or repository. 
However, with the results of Section \textit{Sensitivity Analysis}, the simultaneous sensitivity analysis has demonstrated that our measured causal effect is robust to the strong unobserved confounder.

Although we have conducted qualitative studies and identified social signaling as a possible mechanism of the causal effect of emojis, it is not yet verified whether such signal can be widely accepted by other developers. Such questions would be best answered with a user study, such as a structured interview, a survey, or an randomized controlled experiment, which complements this study from the receiver side of the communication. We may expect the user study to verify whether readers can perceive emojis in issues as social signals, and be encouraged to participate or help resolving the issue.

Emojis are widely considered a ubiquitous language across languages. Our work only focuses on the GitHub English issues, and for the future work, we can expand our experiments to other languages or even other domain-specific communities to make our conclusion more generalizable. 

Results from the causal inference provide new insights into the relationship between issue contents, emoji usage, and reader response on GitHub. We discuss the implications of our work below and hope that our work can motivate future research.

Our research reveals the causality between emoji usage and issue outcomes. We expect that authors can be motivated to use more emojis in issues, and increased efficiency of issue resolution can also attract more developers to the platform. In addition to putting more emojis in the issues, our results in Section \textit{Heterogeneous Effects by Issue Types} also benefit the authors in deciding which emojis to add according to the issue types. Authors may choose emojis that signal the type and content of the issue. For example, if writing an issue about raising bugs, authors can put more emojis with domain-specific semantics, such as \MyEmoji{\stopsign} (stop sign) or \MyEmoji{\bugemoji} (bug) to attract other developers' attention. For workplaces that would like to remain hybrid or remote, our work provides empirical evidence that adopting emojis could be a step to improve working efficiency.

% Moreover, for further studies focusing on building emoji recommender systems on GitHub, our case study and the fine-grained analysis of heterogeneous emoji effect provides the insight that issue types are significant factors to consider in the recommendation process. For example, when encountering a bug issue, the recommender system may consider more relevant emojis for bug reports.
% \blue{wei: Maybe we can remove the last paragraph all together? Seems we have write a lot of implimcations..}

%% file: sections/8_conclusion.tex
\section{Conclusion}
In this work, we present the first empirical study of the causality between emoji usage and GitHub issue outcomes. To this end, we construct a dataset of more than 200K GitHub issues, as well as the data of potential confounders from the issue texts, the issue author, and repository information. We use propensity score matching to quantify causality and multivariate logistic regression to estimate the propensity score. We show significant effects of emoji usage on issues from two perspectives: issue participation and resolution. 
In the fine-grained analysis, we show the heterogeneous treatment effects by issue types, and to understand causality, we also explore HTE on type-associated emojis. More salient treatment effects caused by type-associated emojis preliminarily verify our hypothesis that emojis, as social signals, reduce information asymmetry to promote user participation and issue resolution.
The promising results suggest that developers can add more emojis to help resolve their issues and attract more users to participate in discussions.

% \vspace{-0.5em}
\section{Broader Perspective and Ethical Considerations}
The causal effects of emojis in increasing developer participation and resolving issues suggest that emojis can have a positive impact on improving work outcomes in online workspace, which may motivate researchers and practitioners to evaluate if the conclusion can be generalized to the real-life workspace to help establish more effective remote working. 
Although there is a dramatic increase in emoji usage on GitHub, the use of emojis is still much lower than on social media platforms, such as Twitter. GitHub website can encourage or guide users to use more emojis in issues.
We have thoroughly discussed the positive and negative outcomes of our results in Section \textit{Limitations and Implications}. In our work, the data in the GitHub issue dataset are all from a third-party public project, GHTorrent, and we do not collect or release other new datasets. During data preprocessing, we strictly follow ethical principles and do not attempt to infer the identities of the issue authors.

\section*{Acknowledgements}
We would like to thank anonymous reviewers as well as Vanessa Frias-Martinez and Paiheng Xu for reviewing the paper and for providing helpful comments and suggestions.

%% file: sections/9_appendix.tex
\appendix
% \vspace{-1.5em}
\section*{Appendix}
\section{Identifying Issue Templates and Bots}
\label{sec:templates}

\begin{figure}[t]
    \centering
    \includegraphics[width=\linewidth]{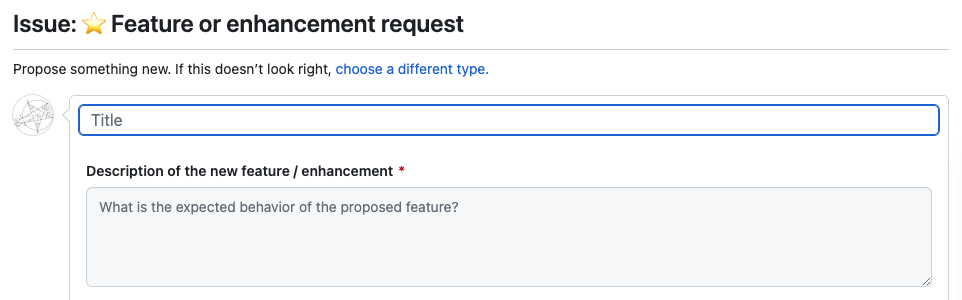}
    % \vspace{-1.5em}
    \caption{An issue template for feature request in the microsoft/PowerToys repository}
    % \vspace{-1em}
    \label{fig:issue_template}
\end{figure}

To formalize the format of GitHub issues, a large number of repository owners set issue templates and force issue authors to use the predefined templates. For example, as shown in Figure \ref{fig:issue_template}, the repository, microsoft/Powertoys,\footnote{\url{https://github.com/microsoft/PowerToys/issues/new/choose}} asks developers to use this template when composing issues related to the feature request. The emoji \MyEmoji{\goldstar} (star) in this template are not added by the issue authors but by the repository owners. We remove all the issues in the template-applied repositories from our dataset by checking whether the repository contains ``ISSUE\_TEMPLATE'' folder.

% Since when setting issue templates in the GitHub repository, the repository will automatically generate a folder called ``ISSUE\_TEMPLATE,'' 
% We use the existence of such a folder to judge whether the repository uses templates. 

\begin{figure*}[ht]
    \centering
    \includegraphics[width=\linewidth]{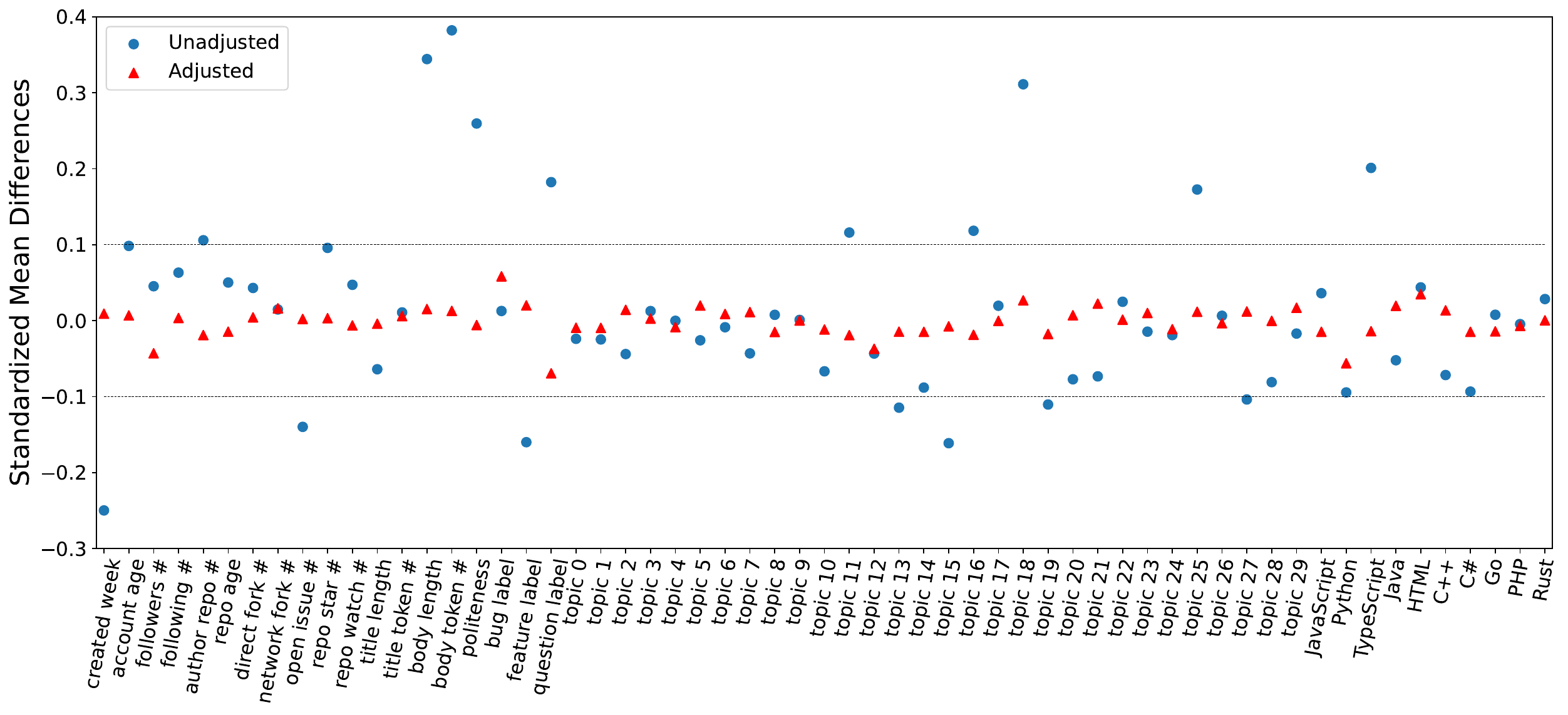}
    % \vspace{-2em}
    \caption{SMD value before matching and after matching for each covariate. After the matching process, the SMD value of all covariates are smaller than 0.1, indicating the similar distribution of all covariates between treatment and control group.}
    % \vspace{-0.5em}
    \label{fig:smd_value}
\end{figure*}

In practical issue usage, a small proportion of repository owners regard issue discussions as their development logs. Owners post their development progress as new issues and close this kind of issues in a short time. Owner-created issues to record progress are not the focus of our project, so we use two criteria to remove them. We first remove the repositories containing more than 10 issues and 90\% issues written by the same author. The second is to remove the issues composed by the repository owners but closed within one day. 

Since GitHub is a code-communication platform, many developers choose GitHub as the first community to test their newly-developed chat bots. The bot-created issues significantly bias the causal effect estimation of emojis. First, we use simhash for near-duplicate detection of issues \cite{manku2007detecting} with the output of an unsigned 64-bit integer, and if there are less than 5 difference digits, two issues are regarded as near-duplicate. We filter the authors with more than 100 issues, and more than 90\% issues of the author are near-duplicate.

\begin{table}[t]
\centering
\small
\begin{tabular}{lrrr}
\toprule
Outcome Variable                   & ATE & Obs. $\Delta$ & Avg. \\ \midrule
\textit{Developer Participation} & & & \\[0.3em]
getting comments                    & 0.075**    & 0.219                    &  0.687 \\
                                         \# of comments                        & 0.101**    & 0.827                    &  2.124 \\
                                         \# of comment users              & 0.162**    & 0.596                    &  1.366 \\
                                          \midrule
\textit{Issue Resolution}                   & & & \\[0.3em] 
\% closed in 180 days & 0.020**    &  0.062                   &  0.500 \\
                                         \% closed in 90 days  & 0.021**
    &  0.055                   &    0.475 \\
                                         \% closed in 60 days  & 0.021**   &  0.054                   &  0.458 \\
                                         \% closed in 30 days  & 0.022**
    &   0.054                  &   0.426  \\
                                         issue closing time (day)          & -1.621**    & -0.079                    &   15.441  \\ \bottomrule
\end{tabular}
% \vspace{-0.5em}
\caption{ATE of using emojis in GitHub issues from July 2018 to June 2019. The meaning of notation is the same as notations in Table \ref{tab:outcome_result}. Signiﬁcance level: ** $p<0.01$, * $p<0.05$, paired $z$-test.}
% \vspace{-1em}
\label{tab:reroduce_outcome}
\end{table}

% \vspace{-0.5em}
\section{Issue Type Classifier}

We extract 160,000 and 20,000 labeled issues as the training and test dataset respectively.
For each issue type, we train a binary Roberta \cite{liu2019roberta} classifier to identify whether the issue belongs to this type. When preparing the training positive instances, we first normalize the issue labels to increase the number of positive data. For example, we treat every label with the ``Bug'' substring (except ``Not-Bug'') as a bug label. The three trained classifiers can achieve 89.3\%, 89.21\% and 87.00\% in the bug, feature, and question test sets, respectively. After automatically annotating with these three classifiers, we obtain three labels to indicate whether the issue is bug-related, question-related, or feature-related.

\section{SMD Balance Check and Refinement}
\label{sec:smd}
SMD is defined as: 
$SMD = \frac{\overline{X_1} - \overline{X_2}}{\sqrt{(S_1^2 + S_2^2) / 2}}$, where $\overline{X_1}$ and $\overline{X_2}$ are the sample mean for the treated and control groups, respectively. $S_1^2$ and $S_1^2$ are sample variances for two groups.

% Specifically, for binary variables, SMD is defined as
% \begin{equation}
%     SMD = \frac{\hat{p_1} - \hat{p_2}}{\sqrt{(\hat{p_1}(1-\hat{p_1}) + \hat{p_2}(1 - \hat{p_2})) / 2}}
% \end{equation}

% where $\hat{p_1}$ and $\hat{p_2}$ are the prevalence of dichotomous variables in the treated and control groups, respectively. 
We calculate the SMD value and the result show that SMD value for confounder ``\# of characters in the issue body'' (issue body length) is greater than 0.1. The SMD value of issue body length before matching is 0.344, which indicates that this confounder distribution has a large gap between treatment and control groups and only matching on the propensity score can not guarantee the balanced distribution. 

To make the confounders more balanced, we make another refinement to the NNM method: the difference of issue body length in the matched pairs should not exceed 1 SD of the issue body length. For the matching procedure, we first filter the untreated issues that exceed the restriction of the propensity score and the body length of the issue and then find the issue with the closest propensity score. The SMD values of each confounder covariate before and after the matching are visualized in Figure \ref{fig:smd_value}. With the refinement, SMD values of all confounders are below 0.1 threshold.

\section{Causal Inference Reproduction}
\label{sec:reproduce_result}
\subsection{Reproduction on a different year}
We re-collect the GitHub issues in public repositories between July 1, 2018 and June 30, 2019 via GHTorrent and sample with a ratio of 1:20. We repeat the data preprocessing and obtain a new dataset with 361,257 issues where 17,630 issues contain one or more emojis. After propensity score estimation and conducting a NNM, we calculate the average treatment and report the results in Table \ref{tab:reroduce_outcome}. ATE values in Table \ref{tab:reroduce_outcome} are still significant and comply with the findings in the causal effect for issues from 2020 to 2021.

\begin{table}[t]
\centering
\small
\resizebox{\columnwidth}{!}{%
\begin{tabular}{lccc}
\toprule
Outcome Variable                   & ATE (45 topics) & ATE (15 topics) & ATE (no politeness) \\ \midrule
\textit{Developer Participation} & & & \\[0.3em]
getting comments                    & 0.041**    & 0.046**                    &  0.045** \\  
\# of comments                        & 0.127**    & 0.098**                    &  0.111**\\
\# of comment users              & 0.117**    & 0.127**                    &  0.122** \\
                                          \midrule
\textit{Issue Resolution}                   & & & \\[0.3em] 
\% closed in 180 days & 0.036** & 0.033** & 0.035** \\
                                         \% closed in 90 days  & 0.035** & 0.031** & 0.034** \\
                                         \% closed in 60 days  & 0.039** & 0.035** & 0.036** \\
                                         \% closed in 30 days  & 0.040** & 0.034** & 0.036** \\
                                         issue closing time (day) & -1.961** & -1.765**& -1.730** \\ \bottomrule
\end{tabular}
}
% \vspace{-0.5em}
\caption{Average treatment effect of using emojis in issues from June 2020 to June 2021 with 45-dimensional topic distribution, with 15-dimensional topic distribution, and without politeness score in propensity score estimation. The meaning of notation is the same as the notation in Table \ref{tab:outcome_result}. Signiﬁcance level: ** $p<0.01$, * $p<0.05$, paired $z$-test.}
% \vspace{-1em}
\label{tab:reproduce_topic}
\end{table}

\subsection{Reproduction on various covariate specifications}

To verify our results’ robustness to misspecifications of the covariates, we repeat our analyses on three alternative specifications. In the first two alternatives, we set the numbers of topics as 45 and 15, re-train the LDA model, and repeat the analyses. In the third alternative specification, we remove the politeness score from the propensity score estimation and show all the values in Table \ref{tab:reproduce_topic}. We observe that the estimated treatment effects under the three alternative specifications all share a similar size and significance level as our main result, which indicates that our findings are robust to misspecification in estimating the covariates.

\subsection{Reproduction on GBRT model}

To test our findings’ sensitivity to the model of propensity score estimation, we repeat the analysis by using the Gradient Boosted Regression Tree (GBRT) model for propensity score estimation. We report the estimates in Table \ref{tab:reproduction_gbrt}. The treatment effects are similar to our main result in effect size and significance level, which suggests the robustness of our estimation to the model of propensity score estimation.

\begin{table}[t]
\centering
\small
\begin{tabular}{lr}
\toprule
Outcome Variable                   & ATE \\ \midrule
\textit{Developer Participation} & \\[0.3em]
getting comments                    & 0.033**    \\
                                         \# of comments                        & 0.094**    \\
                                         \# of comment users              & 0.089**    \\
                                          \midrule
\textit{Issue Resolution}                   & \\[0.3em] 
\% closed in 180 days & 0.022**     \\
                                         \% closed in 90 days  & 0.023**
     \\
                                         \% closed in 60 days  & 0.025**   \\
                                         \% closed in 30 days  & 0.025**
    \\
                                         issue closing time (day)          & -2.556**   \\ \bottomrule
\end{tabular}
% % \vspace{-0.5em}
\caption{Average treatment effect of using emojis in GitHub issues from June 2020 to June 2021 with GBRT model as propensity score estimation method. Signiﬁcance level: ** $p<0.01$, * $p<0.05$, paired $z$-test.}
% % \vspace{-1.5em}
\label{tab:reproduction_gbrt}
\end{table}